\documentstyle[12pt,a4wide]{article}
\input epsf

\newcommand{\news}{\setcounter{equation}{0} \ \indent}
\newcommand{\be}{\begin{equation}}
\newcommand{\ee}{\end{equation}}
\newcommand{\bea}{\begin{eqnarray}}
\newcommand{\eea}{\end{eqnarray}}
\newcommand{\bean}{\begin{eqnarray*}}
\newcommand{\eean}{\end{eqnarray*}}
\font\upright=cmu10 scaled\magstep1
\font\sans=cmss12
\newcommand{\ssf}{\sans}
\newcommand{\stroke}{\vrule height8pt width0.4pt depth-0.1pt}
\newcommand{\Z}{\hbox{\upright\rlap{\ssf Z}\kern 2.7pt {\ssf Z}}}

\newcommand{\C}{{\rlap{\rlap{C}\kern 3.8pt\stroke}\phantom{C}}}
\newcommand{\R}{\hbox{\upright\rlap{I}\kern 1.7pt R}}
\newcommand{\CP}{\C{\upright\rlap{I}\kern 1.7pt P}}

\newcommand{\half}{\frac{1}{2}}

\begin{document}
\pagestyle{plain}
\title{
\begin{flushright}
{\normalsize DAMTP 95-39} \\
\end{flushright}
\vskip 20pt
{\bf The moduli space metric for tetrahedrally 
symmetric 4-monopoles} \vskip 20pt}
\author{Paul M. Sutcliffe\thanks{
Address from September 1995,
 Institute of Mathematics,
University of Kent at Canterbury, Canterbury CT2 7NZ.
 Email P.M.Sutcliffe@ukc.ac.uk
} \\[20pt]
{\sl Department of Applied Mathematics and Theoretical Physics} \\[5pt]
{\sl University of Cambridge} \\[5pt]
{\sl Cambridge CB3 9EW, England} \\[20pt]
{\sl email\  p.m.sutcliffe@damtp.cam.ac.uk}\\[10pt]}

\date{July 1995\\[20pt]
{\small To appear in Physics Letters B}\\[10pt]
}

\maketitle

\begin{abstract}
The metric on the moduli space of SU(2) charge four BPS monopoles 
with tetrahedral symmetry is calculated using numerical methods.
In the asymptotic region, in which the four monopoles are
located on the vertices of a large tetrahedron, the metric is
in excellent agreement with the point particle metric.
We find that the four monopoles are accelerated through the cubic monopole
configuration and compute the time advance.
Numerical evidence is presented for a remarkable equivalence
between a proper distance 
in the 4-monopole moduli space and a related proper distance
in the point particle moduli space.
This equivalence implies that the approximation to the 
time advance (and WKB quantum phase
shift) calculated using the point particle derived metric is 
exact.

\end{abstract}
\newpage
\section{Introduction}
\news
The dynamics of SU(2) BPS monopoles may be approximated
by the time evolution of a finite number of collective
coordinates \cite{M,S}. In this moduli space approach the dynamics of
$k$ monopoles is approximated by geodesic motion on the $k$-monopole
moduli space ${\cal M}_k$, which is a $4k$-dimensional manifold. To study 
monopole dynamics therefore requires the construction of the metric
on ${\cal M}_k$, which is determined by the kinetic part of the 
field theory action. In the case $k=2$, Atiyah and Hitchin \cite{AH}
were able to calculate the metric using indirect methods and making
use of its hyperk\"{a}hler property. However, for $k>2$ the problem 
is a more difficult one and no metrics have yet been calculated.
Recently, Gibbons and Manton 
\cite{GM} have calculated, for general $k$, the asymptotic metric
on regions of ${\cal M}_k$ which describe well-separated monopoles.
This asymptotic metric is obtained by treating the monopoles as point
particles and is of a generalized Taub-NUT form.

The moduli space ${\cal N}$, of tetrahedrally symmetric 4-monopoles, is a
one-dimensional totally geodesic submanifold of ${\cal M}_4$,
and the associated four-monopole scattering process has been 
investigated in some detail \cite{HSa}. 
The fact that ${\cal N}$ is one-dimensional allows the 
monopole trajectories
to be determined even though the metric is not
known. In this paper we construct the metric on ${\cal N}$ 
by working with Nahm data and using numerical methods. 
In the asymptotic region, in which the four monopoles are
located on the vertices of a large tetrahedron, the metric is
in excellent agreement with the point particle metric.
We use this metric to calculate the time advance/delay.
We also provide numerical evidence for the following rather remarkable 
equivalence.
Let $l$ be a good global coordinate on ${\cal N}$, such that
$l=0$ represents coincident monopoles and large $l$ represents
well-separated monopoles. Then the proper distance from the point
$l=0$ to a point with $l$ large is equal to the proper distance
from the singularity to the same point $l$ in the generalized Taub-NUT
space. 
This equivalence implies that the approximation to the 
time advance (and quantum phase
shift) calculated using the point particle derived metric is 
exact.
This is similar to 
a previous numerical result found in the 2-monopole case \cite{B},
and suggests that it may be a general feature of the point particle
metric.

\section{Four monopoles with tetrahedral symmetry}
\news
In this section we recall the results \cite{HSa} on tetrahedrally symmetric
charge four monopoles that we shall require later.
Monopoles are equivalent to various other kinds of mathematical
creatures, and here we shall use two of these; namely, spectral curves
and Nahm data.
Spectral curves \cite{Ha} are algebraic curves in the holomorphic
tangent bundle to the Riemann sphere. 
Let $\zeta$ be the standard inhomogeneous coordinate on the base space
and $\eta$ the fibre coordinate. Then a 4-monopole with tetrahedral
symmetry has a spectral curve
\be
\eta^4+ i36a\kappa^3\eta\zeta(\zeta^4-1)+
3\kappa^4(\zeta^8+14\zeta^4+1)=0.\label{sc}
\ee
where $a\in(-a_c,a_c)$, with $a_c=3^{-5/4}\sqrt{2}$,
and $\kappa$ is half the real period of 
the elliptic curve
\be
y^2=4(x^3-x+3a^2).
\ee
Hence there is a one-parameter family of tetrahedrally symmetric
4-monopoles. Since this family of monopoles is singled out from the
general 4-monopole configuration by the imposition of a symmetry, this 
implies that the corresponding submanifold ${\cal N}\subset{\cal M}_4$
is totally geodesic. The associated monopole dynamics 
has been studied in detail and describes the scattering of four 
monopoles which are initially well-separated and positioned on the
vertices of a contracting regular tetrahedron. As the monopoles
merge they scatter instantaneously through a configuration
with cubic symmetry and emerge on the vertices of an expanding
tetrahedron dual to the incoming one.
For the purposes of constructing the metric on ${\cal N}$ we
require a good global coordinate which we can identify with the
distance of each monopole from the origin, at least when the monopoles
are well-separated.

Let ${\bf x}_1,{\bf x}_2,{\bf x}_3,{\bf x}_4$ be the four points,
each a distance $\vert l\vert$ from the origin, given by
\bea
&{\bf x}_1=(-l,-l,-l)\frac{1}{\sqrt{3}} \nonumber\\
&{\bf x}_2=(-l,+l,+l)\frac{1}{\sqrt{3}} \nonumber\\
&{\bf x}_3=(+l,+l,-l)\frac{1}{\sqrt{3}} \label{vt}\\
&{\bf x}_4=(+l,-l,+l)\frac{1}{\sqrt{3}}. \nonumber
\eea
They are the vertices of the tetrahedron on which the monopoles 
are located when they are well-separated.
For well-separated monopoles the asymptotic spectral curve
can be obtained as a product of the individual monopole's
spectral curves. If $\vert l\vert$ is large and we take the four 
monopoles to have positions ${\bf x}_1,{\bf x}_2,{\bf x}_3,{\bf x}_4$, 
then we obtain the asymptotic spectral curve
\be
\eta^4+ i\frac{16}{3^{3/2}}l^3\eta\zeta(\zeta^4-1)+
\frac{4}{9}l^4(\zeta^8+14\zeta^4+1)=0.\label{sca}
\ee
The spectral curve (\ref{sc}) has this form in the limit 
$a\rightarrow a_c$, upon which $\kappa\rightarrow\infty$.
By comparing  (\ref{sc}) and (\ref{sca}) we see that 
we can make the identification 
\be
l=\Lambda a^{1/3}\kappa, \hskip 1cm \mbox{where } 
\hskip 1cm \Lambda=3^{7/6}2^{-2/3}.
\ee
At $a=0$ (\ref{sc}) is the spectral curve of the cubic 4-monopole
\cite{HMM,HSa}, which has all four Higgs zeros at the origin.
If we define, in the usual way, the positions of the monopoles to be
given by the zeros of the Higgs field, then $l=0$ is when all four
monopoles have zero distance from the origin. Hence $l\in$\ \R\ is a good
global coordinate on ${\cal N}$ with a natural interpretation
as the distance of each monopole from the origin.

We have used the spectral curve approach to  monopoles to identify
a convenient coordinate on ${\cal N}$, but in order to discuss the
metric we now need to turn to the ADHMN formulation \cite{N,Hb}.
This is usually presented in terms of Nahm data consisting of 
three Nahm matrices $(T_1,T_2,T_3)$, but in order to discuss the
metric we must, following Donaldson \cite{D}, introduce a fourth
Nahm matrix $T_0$. Then we have that charge $k$ monopoles are 
equivalent to Nahm data $(T_0,T_1,T_2,T_3)$, 
which are four $k\times k$ matrices which depend
on a real parameter $s\in[0,2]$ and satisfy the following;\\
\newcounter{con}
\setcounter{con}{1}
(\roman{con})  Nahm's equation
\be
\frac{dT_i}{ds}+[T_0,T_i]=
\half\epsilon_{ijk}[T_j,T_k] \hskip 30pt i=1,2,3\nonumber
\ee\\

\addtocounter{con}{1}
(\roman{con}) $T_0$ is regular for $s\in[0,2]$.
$T_i(s)$, $i=1,2,3$, is regular for $s\in(0,2)$ and has simple
poles at $s=0$ and $s=2$,\\

\addtocounter{con}{1}
(\roman{con}) the matrix residues of $(T_1,T_2,T_3)$ at each
pole form the irreducible $k$-dimensional representation of SU(2),\\

\addtocounter{con}{1}
(\roman{con}) $T_i(s)=-T_i^\dagger(s)$, \hskip 30pt $i=0,1,2,3$,\\

\addtocounter{con}{1}
(\roman{con}) $T_i(s)=T_i^t(2-s)$, \hskip 30pt $i=0,1,2,3$.\\

\setcounter{con}{5}

Let $G$ be the group of analytic $su(k)$-valued functions $h(s)$,
for $s\in[0,2]$, which are the identity at $s=0$ and $s=2$, and 
satisfy $h^t(2-s)=h^{-1}(s)$.
Then gauge transformations $h\in G$ act on Nahm data as
\be
T_0\rightarrow hT_0h^{-1}-\frac{dh}{ds}h^{-1} \hskip 20pt\\
\ee
\be
T_i\rightarrow hT_ih^{-1}  \hskip 15pt i=1,2,3. \nonumber
\ee

Note that the gauge $T_0=0$ may always be chosen, which is why 
this fourth Nahm matrix is usually not introduced. However, when
discussing the metric on Nahm data we need to consider the action of
the gauge group and so this extra Nahm matrix needs to be kept, at
least temporarily.

In the gauge $T_0=0$ the Nahm data corresponding to a tetrahedrally
symmetric 4-monopole, whose spectral curve we have discussed above,
is given by \cite{HSa}
\be
T_i(s)=x(s)X_i+y(s)Y_i+z(s)Z_i \hskip 20pt i=1,2,3
\label{tnd}
\ee
where $x,y,z$ are the real functions
\begin{eqnarray} x(s)&=&\frac{\kappa}{5}\left(-2\sqrt{\wp(\kappa
    s)}+\frac{1}{4}\frac{\wp^\prime(\kappa s)}{\wp(\kappa
    s)}\right)\\
y(s)&=&\frac{\kappa}{20}\left(\sqrt{\wp(\kappa
    s)}+\frac{1}{2}\frac{\wp^\prime(\kappa s)}{\wp(\kappa s)}\right)\\
z(s)&=&\frac{a\kappa}{2\wp(\kappa s)}.
\label{xyz}\end{eqnarray}
Here $\wp$ is the Weierstrass function satisfying
\be \wp^{\prime 2}=4\wp^3-4\wp+12a^2
\label{wfun}\ee
with prime denoting differentiation with respect to the argument.
The tetrahedrally symmetric Nahm triplets are
{\small
$$
(X_1,X_2,X_3)=\left(
\left[{\begin{array}{cccc}
0&\sqrt{3}&0&0\\
-\sqrt{3}&0&2&0\\
0&-2&0&\sqrt{3}\\
0&0&-\sqrt{3}&0 \end{array}}\right],
\left[{\begin{array}{cccc}
0&i\sqrt{3}&0&0\\
i\sqrt{3}&0&2i&0\\
0&2i&0&i\sqrt{3}\\
0&0&i\sqrt{3}&0 \end{array}}\right],
\left[{\begin{array}{cccc}
3i&0&0&0\\
0&i&0&0\\
0&0&-i&0\\
0&0&0&-3i \end{array}}\right]
\right)
$$
$$
(Y_1,Y_2,Y_3)=
2
\left(
\left[{\begin{array}{cccc}
0&-\sqrt{3}&0&-5\\
\sqrt{3}&0&3&0\\
0&-3&0&-\sqrt{3}\\
5&0&\sqrt{3}&0 \end{array}}\right],
\left[{\begin{array}{cccc}
0&-i\sqrt{3}&0&5i\\
-\sqrt{3}i&0&3i&0\\
0&3i&0&-\sqrt{3}i\\
5i&0&-\sqrt{3}i&0 \end{array}}\right],
\left[{\begin{array}{cccc}
2i&0&0&0\\
0&-6i&0&0\\
0&0&6i&0\\
0&0&0&-2i \end{array}}\right]
\right)$$
$$
(Z_1,Z_2,Z_3)=\sqrt{3}\left(
\left[{\begin{array}{cccc}
0&i&0&0\\
i&0&0&0\\
0&0&0&-i\\
0&0&-i&0 \end{array}}\right],
\left[{\begin{array}{cccc}
0&1&0&0\\
-1&0&0&0\\
0&0&0&-1\\
0&0&1&0 \end{array}}\right],
\left[{\begin{array}{cccc}
0&0&1&0\\
0&0&0&1\\
-1&0&0&0\\
0&-1&0&0 \end{array}}\right]
\right)$$
}
Although this Nahm data does not satisfy property (\roman{con})
this can be achieved by a suitable change of basis.

In the next section we shall use this Nahm data to calculate the
metric on ${\cal N}$.

\section{Calculation of the metric}
\news
It is known that the transformation between the  monopole 
moduli space metric and the metric on Nahm data is an 
isometry \cite{H,NA}. We can therefore calculate the metric, $g(l)$,
on ${\cal N}$ by computing the metric on the Nahm data given in the 
previous section. This requires the computation of the tangent vector
$(V_0,V_1,V_2,V_3)$ corresponding to the point 
with Nahm data $(T_0,T_1,T_2,T_3)$. In principal,
since the Nahm data is explicitly known,  this could be 
achieved by direct differentiation,
\be V_i=\frac{dT_i}{dl}. \label{dd} \ee
However, this is not a practical way to proceed since the
Weierstrass function (\ref{wfun}), in terms of which the Nahm data
is given, itself depends on $l$
(through its dependence on $a$). Instead we calculate the tangent
space to ${\cal N}$ by solving the linearized Nahm equation
\be
\dot V_i +[V_0,T_i]+[T_0,V_i]=\epsilon_{ijk}[T_j,V_k]
\hskip 20pt i=1,2,3
\label{lne}
\ee
and
\be
\dot V_0+\sum_{i=0}^3 [T_i,V_i]=0
\label{bg}
\ee
where $V_i,\ i=0,1,2,3$, is an analytic $su(4)$-valued function
of $s\in[0,2]$. Dot denotes differentiation with respect to $s$.
The metric on Nahm data is then given by
\be
g(l)=-\Omega\int_0^2 \sum_{i=0}^4 \mbox{tr}(V_i^2)\ ds
\label{mnd}
\ee
where tr denotes trace and $\Omega$ is a normalization constant.

From now on we use the gauge freedom to set $T_0=0$. Equation
(\ref{bg}) is the background gauge constraint which ensures that
the tangent vectors we compute are horizontal {\sl ie} that the
tangent vectors are orthogonal to the gauge orbits. The tangent
vectors are tetrahedrally symmetric so we may write
\be
V_i=q_1X_i+q_2Y_i+q_3Z_i \hskip 20pt i=1,2,3
\label{ttv}
\ee
where ${\bf q}=(q_1,q_2,q_3)^t$ is an analytic real 3-vector 
function of $s\in[0,2]$. It is easily checked that the tetrahedral
symmetry of the Nahm triplets implies the following identities
\be
\sum_{i=1}^3[X_i,Y_i]=\sum_{i=1}^3[X_i,Z_i]=\sum_{i=1}^3[Y_i,Z_i]=0.
\label{iden}
\ee
Substituting (\ref{tnd}) and (\ref{ttv}) into 
the background gauge equation (\ref{bg}) and using the identities
(\ref{iden}) gives the solution $V_0=0$. The remaining equations
(\ref{lne}) become the following equation for the 3-vector ${\bf q}$
\be
\dot{\bf q}=M{\bf q} \hskip 20pt \mbox{where} \hskip 10pt
M=\left[{\begin{array}{ccc}
4x&-96y&-12z/5\\
-6y&-16y-6x&-6z/5\\
-4z&-32z&-4x-32y
\end{array}}\right].
\label{ode}
\ee
Substituting (\ref{ttv}) into (\ref{mnd}) gives the metric in terms of
${\bf q}$ as
\be
g=12\Omega \int_0^2 (5q_1^2+80q_2^2+3q_3^2) \ ds.
\label{met}
\ee

The ordinary differential equation (\ref{ode}) has regular-singular 
points at $s=0$ and $s=2$, since the functions appearing in $M$ have first order
poles at these points. Analysis of the initial value problem
at $s=0$ reveals that there is a two-dimensional family of solutions
to (\ref{ode}) which are normalizable for $s\in[0,2)$. They are given
by the two-parameter, $\alpha_1,\alpha_2$, family of initial
conditions
\be
{\bf q}\sim(0,\alpha_1 s^3,\alpha_2 s^2)^t \hskip 10pt 
\mbox{as} \hskip 10pt s\sim 0.
\label{icl}
\ee
Repeating the analysis for the initial value problem at $s=2$ gives
a two-parameter, $\beta_1,\beta_2$, family of normalizable
solutions for $s\in(0,2]$, with initial conditions
\be
{\bf q}\sim(16\beta_1(2-s)^3,3\beta_1(2-s)^3,\beta_2(2-s)^2)^t \hskip 10pt 
\mbox{as} \hskip 10pt s\sim 2.
\label{icr}
\ee

The solution we require is the one-parameter family  which is 
normalizable in the closed interval $s\in[0,2]$. For later
convenience we take $\alpha_2$ to be the free parameter which describes
this family of solutions. To compute these solutions
by solving an initial value problem would be a difficult shooting 
problem if it were not for the fact that equation (\ref{ode}) 
is linear. This reduces the task to a simple problem in linear algebra
which we implement as follows. Given a value for $\alpha_2$, say
$\alpha$, let ${\bf p}_1(s)$
denote the solution ${\bf q}(s)$ of (\ref{ode}) corresponding to the
initial conditions (\ref{icl}) with $(\alpha_1,\alpha_2)=(0,\alpha)$.
This solution is calculated for $s\in[0,1]$. Numerically we compute
this solution using a fourth order Runge-Kutta method. 
Let ${\bf p}_2(s)$ denote a second solution, but this time with 
initial conditions $(\alpha_1,\alpha_2)=(1,0)$. Similarly, let
${\bf p}_3(s)$ and ${\bf p}_4(s)$ be the solutions calculated for
$s\in[1,2]$ obtained from the initial conditions (\ref{icr}) with
parameter values
$(\beta_1,\beta_2)=(1,0)$ and $(\beta_1,\beta_2)=(0,1)$ respectively.
Next form the $3\times 4$ matrix
\be
U=\left[ 
\begin{array}{cccc}
|&|&|&|\\
{\bf p}_1(1)& {\bf p}_2(1)&{\bf p}_3(1) & {\bf p}_{4}(1) \\
|&|&|&|
\end{array}\right]
\label{umatrix}
\ee
and find the unique solution of the linear matrix equation
\be
U{\bf w}={\bf 0}
\label{keru}
\ee
for ${\bf w}=(1,w_2,w_3,w_4)^t$.
Numerically this is performed by row reduction of the matrix $U$
followed by back substitution.
Then the required solution ${\bf q}(s)$ is given by
 \be
{\bf q}(s)=\left\{
\begin{array}{ll}
\ {\bf p}_1(s)+w_2{\bf p}_2(s) & \mbox{if $0\leq s\leq
1$}\\
 & \\
-w_3{\bf p}_3(s)-w_4{\bf p}_4(s) & \mbox{if $1< s\leq
2$}
\end{array}
\right.
\ee
To summarize, the above procedure consists in integrating (\ref{ode}) 
twice from each end of the interval $[0,2]$ to the centre
and then finding a linear combination of these solutions which
match at the centre.

We note that in the special case $l=0$ ({\sl ie} $a=0$), which
corresponds to the cubic monopole, the tangent vector may be
calculated explicitly in closed form. In this case the third component
of ${\bf q}$ decouples from the other two and we have the solution
${\bf q}=(0,0,q_3)^t$ with
\be
q_3=\frac{\alpha_2}{\kappa^2\wp(\kappa s)}
\ee
where $\kappa$ and $\wp$ take their values corresponding to $a=0$.

The next issue we confront is to ensure that the tangent vector
we compute is dual to the coordinate $l$. This requires the
determination of the correct $l$-dependent normalization factor
$\alpha_2$. In terms of ${\bf q}$ the equation (\ref{dd}) becomes
$(q_1,q_2,q_3)=(dx/dl,dy/dl,dz/dl)$. We calculate the correct
normalization factor by considering the third component,
$q_3={dz/dl}$,
 in the limit $s\rightarrow 0$. 
Substituting the asymptotic behaviour of $\wp(\kappa s)$ as $s\sim 0$
into the expression (\ref{xyz}) for $z$ and comparing with the 
definition of $\alpha_2$ given by (\ref{icl}) we obtain
$$\alpha_2 s^2=\frac{d}{dl}(\frac{1}{2}a\kappa^3 s^2)$$
which gives
\be
\alpha_2=\frac{3l^2}{2\Lambda^3}.
\ee

There is now only one constant left to determine, which is the 
overall metric factor $\Omega$. This is fixed by the requirement
that the metric tends to the sum of the monopole masses in the limit
of infinite separation {\sl ie} $g(l)\rightarrow 16\pi$ as 
$l\rightarrow\infty$.

We now apply the above numerical scheme to calculate the metric.
The integral in (\ref{met}) is calculated using a standard composite
Simpsons rule. The result is displayed in Fig. 1 (solid curve) 
for $0\leq l \leq 18$. We see that the metric is a monotonic
increasing function of $l$, which implies that the monopoles speed
up as they approach each other and are accelerated through the cubic 
monopole configuration $l=0$.

Recently, Gibbons and Manton 
\cite{GM} have calculated the asymptotic metric,
on regions of ${\cal M}_k$ which describe well-separated monopoles,
by treating the monopoles as point particles. 
In our case of interest the imposed tetrahedral symmetry implies
that all four monopoles have the same internal phase. This implies
that there are no electric charge differences between the monopoles
so that we can consider pure monopoles not dyons. For $k$ 
well-separated monopoles
with positions ${\bf x}_i$,\  $i=1,..,k$, the point particle lagrangian is
\cite{GM}
\be
{\cal L}=2\pi\sum_{i=1}^k{\dot{\bf x}_i}^2-2\pi\sum_{1\leq i<j\leq k}
\frac{(\dot{\bf x}_i-\dot{\bf x}_j)^2}{\vert {\bf x}_i-{\bf x}_j\vert}.
\ee
For our case of interest $k=4$ and the positions are 
the vertices of the tetrahedron given in equation (\ref{vt}).
Then the above lagrangian becomes
\be
{\cal L}=\frac{1}{2}\widetilde g(l) \dot l^2
\hskip 10pt \mbox{where} \hskip 10pt \widetilde g(l)=
16\pi(1-\frac{\sqrt{6}}{l}).
\label{am}
\ee

The point particle metric $\widetilde g(l)$ is the approximation
to the true metric $g(l)$, and the two should agree in the large $l$
limit. In Fig. 1 we plot the point particle metric (dashed curve) for
comparison with the true metric. We see that indeed the two metrics
are in excellent agreement in the asymptotic limit. This is a useful
check not only on the numerics used in this paper but also on the point
particle approximation applied in \cite{GM}. Note that the metric 
$\widetilde g$ has a singularity ({\sl ie} a point at which its 
determinant changes sign) at $l=\sqrt{6}$.

Given monopoles which are initially well-separated with $l=L$,
the time taken, $t$, for the monopoles to scatter and reach
this separation again may be computed as
$$t=\Delta\sqrt{\frac{2}{\cal T}}$$
where ${\cal T}$ is the total kinetic energy in the system
and $\Delta$ is the proper distance from $L$ to the origin
\be
\Delta=\int_0^{L}\sqrt{g(l)} \ dl.
\label{pd}
\ee
The time advance $\delta t$, due to the acceleration of the monopoles,
 is related to the proper distance via
\be
 \delta t=\sqrt{\frac{2}{\cal T}}(4L\sqrt{\pi}-\Delta).
\ee
Note that the asymptotic behaviour of the metric (which is the point
particle metric (\ref{am})) means that the time advance has a 
logarithmic divergence in the limit $L\rightarrow\infty$.

It is interesting to compare the true time advance with that
obtained in the point particle approximation. In the case of 
2-monopoles the point particle metric is the Taub-NUT metric 
with a negative mass parameter \cite{Mb}. This also has a singularity,
at a finite value $r_0$ of the radial distance $r$ between the monopoles.
Given a point in ${\cal M}_2$ corresponding to large $r$, the 
Atiyah-Hitchin metric can be used to calculate the proper
distance of this point to the bolt (which describes coincident
monopoles). The proper distance from the same point $r$ in 
Taub-NUT space to the singularity can also be calculated and the two
compared. It is curious that  numerically these two  
distances are found to agree \cite{B}. This equivalence,
for which there is at present no explanation, implies that the
approximation to the time advance calculated using the Taub-NUT metric
is exact.

Given that we have the metric $g$ and its point particle 
approximation $\widetilde g$ we can investigate the possibility
that a similar result to that above exists in the 4-monopole case.
The proper distance from the singularity in the generalized Taub-NUT
space is
\bea
\widetilde\Delta&=&\int_{\sqrt{6}}^L\sqrt{\widetilde g(l)}\ dl 
=4\sqrt{\pi}(\sqrt{L(L-\sqrt{6})}+\sqrt{\frac{3}{2}}
\log(\frac{\sqrt{L}-\sqrt{L-\sqrt{6}}}{\sqrt{L}+\sqrt{L-\sqrt{6}}}))
\label{pdgtn}\\
&\sim&4\sqrt{\pi}(L-\sqrt{\frac{3}{2}}(\log(\frac{4L}{\sqrt{6}})+1)
\hskip 10pt \mbox{as} \hskip 10pt L\rightarrow\infty. \nonumber
\eea
It is this which we wish to compare with $\Delta$ in the large $L$ 
limit. Numerically we have computed the metric $g(l)$ at values
up to $l=18$. It can be seen from Fig. 1 that this is a reasonably
large value since the metrics are very similar for $l>10$. 
Setting $L=18$ in (\ref{pdgtn}) gives the result
 $\widetilde\Delta=89.7$.
The integration to calculate $\Delta$ at $L=18$ from the 
numerical values for $\sqrt{g}$ is performed with the
use of the numerical routines {\small FITPACK}. We fit a spline 
under tension to the data values and integrate the resulting spline
to obtain the result $\Delta=89.2$. So, to within the numerical
accuracy of the calculation, we find that the two answers agree.
This implies that there is no relative WKB quantum phase shift from
the point particle approximation
in the quantized dynamics of the above classical motion.

The result in the two monopole case and the numerical evidence 
presented here suggests that perhaps this feature of the point
particle metric is a general one. At present there is little 
explanation for this possible equivalence, but perhaps the answer
lies in some global geometrical properties of the metrics
and the fact that the point particle metric inherits the 
hyperk\"{a}hler  property of the true metric on ${\cal M}_k$.

\section{Conclusion}
We have introduced a numerical scheme to calculate the monopole
moduli space metric from Nahm data. This scheme has been used
to calculate the metric on a totally geodesic submanifold
of the 4-monopole moduli space, corresponding to tetrahedrally
symmetric monopoles. The results compare well with the 
asymptotic point particle metric and we have presented 
evidence for a curious exact result using an approximate metric.
The scheme can be applied to calculate the metric on submanifolds
of ${\cal M}_k$ for which the Nahm data is known. A suitable 
candidate is the submanifold of ${\cal M}_3$ obtained by
imposing a twisted line symmetry on three monopoles \cite{HSc}.
In this scattering process it appears that the zeros of the Higgs
field stick at the origin for a finite time interval. A calculation
of the metric would reveal the time scale over which this sticking
takes place.\\

\ \vskip 20pt

\noindent{\bf Acknowledgements}

Many thanks to Andrew Dancer, Conor Houghton, Robert Leese and
Nick Manton for useful discussions. 
I thank the EPSRC
for a research fellowship.

 \newpage
\begin{figure}[ht]
\begin{center}
\vskip 1cm
\leavevmode
{\epsfxsize=12cm \epsffile{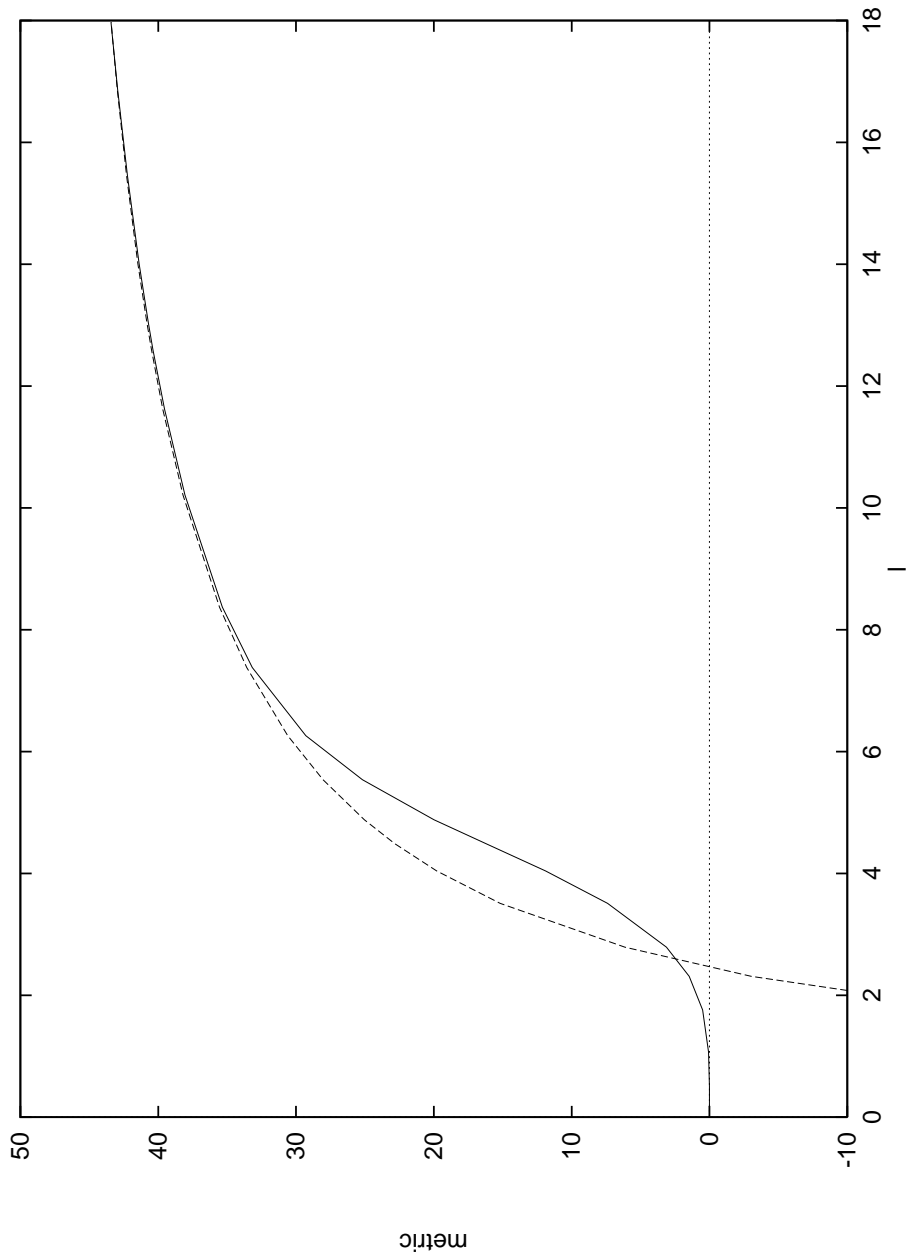}}
\caption{The metric $g(l)$ (solid curve) and the point particle
metric $\widetilde g(l)$ (dashed curve).}
\end{center}
\end{figure}


\begin{thebibliography}{99}
\bibitem{M} N.S. Manton, Phys. Lett. 110B, 54 (1982).
\bibitem{S} D. Stuart, Commun. Math. Phys. 166, 149 (1994).
\bibitem{AH} M.F. Atiyah and N.J. Hitchin,
\lq{\sl The geometry and dynamics of magnetic monopoles}\rq,
Princeton University Press, 1988.
\bibitem{GM} G.W. Gibbons and N.S. Manton, \lq{\sl The moduli
space metric for well-separated BPS monopoles}\rq, Cambridge
preprint DAMTP 95-29.
\bibitem{HSa} C.J. Houghton and P.M. Sutcliffe, \lq{\sl Tetrahedral and
    cubic monopoles}\rq, Cambridge preprint DAMTP 95-13.
\bibitem{B} B.J. Schroers, Nucl. Phys. B367, 177 (1991).
\bibitem{Ha} N.J. Hitchin, Commun. Math. Phys. 83, 579 (1982). 
\bibitem{HMM} N.J. Hitchin, N.S. Manton and M.K. Murray,
\lq{\sl Symmetric Monopoles}\rq, Cambridge preprint DAMTP 95-17.
\bibitem{N} W. Nahm, \lq{\sl The construction of all self-dual
multimonopoles by the ADHM method}\rq, in Monopoles in quantum field
theory, eds. N.S. Craigie, P. Goddard and W. Nahm, World Scientific,
1982.
\bibitem{Hb} N.J. Hitchin, Commun. Math. Phys. 89, 145 (1983).
\bibitem{D} S.K. Donaldson, Commun. Math. Phys. 96, 387 (1984).
\bibitem{H} J. Hurtubise, Abstracts. Am. Math. Soc. 11, 268 (1990).
\bibitem{NA} H. Nakajima, \lq{\sl Monopoles and Nahm's equations}\rq,
University of Tokyo preprint, 1991.
\bibitem{Mb} N.S. Manton, Phys. Lett. 154B, 397 (1985); (E) 157B 475
(1985).
\bibitem{HSc} C.J. Houghton and P.M. Sutcliffe, \lq{\sl Monopole
   scattering with a twist}\rq, Cambridge preprint DAMTP 95-28.
\end{thebibliography}
\end{document}